\documentclass[%
longbibliography,
 amsmath,amssymb,
 aps,
twocolumn,
]{revtex4-1}

\usepackage{graphicx}
\usepackage{dcolumn}
\usepackage{bm}
\usepackage{upgreek}
\usepackage{hyperref}
\DeclareMathAlphabet{\mathpzc}{OT1}{pzc}{m}{it}

\newcommand{\m}{\mathrm}

\newcommand{\fref}[1]{Fig.~\ref{#1}}

\begin{document}

\title{Interfacing planar superconducting qubits with high overtone bulk acoustic phonons}

\author{Mikael Kervinen}
\author{Ilkka Rissanen}
\author{Mika Sillanp\"a\"a}%
 \email{Mika.Sillanpaa@aalto.fi}
\affiliation{%
 Department of Applied Physics, Aalto University, P.O. Box 15100, FI-00076 AALTO, Finland
}%


\date{\today}

\begin{abstract}

Mechanical resonators are a promising way for interfacing qubits in order to realize hybrid quantum systems that offer  great possibilities for applications. Mechanical systems can have very long energy lifetimes, and they can be further interfaced to other systems. Moreover, integration of mechanical oscillator with  qubits creates a potential platform for exploration of quantum physics in macroscopic mechanical degrees of freedom. Utilization of high overtone bulk acoustic resonators coupled to superconducting qubits is an intriguing platform towards these goals. These resonators exhibit a combination of high frequency and high quality factors. They can reach their quantum ground state at dilution refrigeration temperatures and they can be strongly coupled to superconducting qubits via their piezoelectric effect. In this report, we demonstrate our system where bulk acoustic phonons of a high overtone resonator are coupled to a transmon  qubit in a planar circuit architecture. We show that the bulk acoustic phonons are interacting with the qubit in the simple design architecture at the quantum level, representing further progress towards quantum control of mechanical motion.




\end{abstract}

\maketitle


\section{Introduction}

Superconducting circuits are at the forefront of the field of quantum information processing and they are among the leading platforms for realizing quantum computing. Superconducting qubits are well suited for their specific tasks with wide range of applications \cite{WendinReview2017}. However, it will be beneficial to extend their properties by using hybrid systems involving disparate degrees of freedom, so that it is possible to take advantage of the complementary functionalities of different quantum systems. Superconducting qubits have been experimentally integrated with for instance spin ensembles \cite{SembaQBspin,EsteveSpin2011} and magnons \cite{Tabuchi2015}.


Another potential hybrid platform for superconducting or other qubits are micromechanical systems. Due to their potentially simple physical structure, somewhat macroscopic mechanical oscillators can exhibit ultrasmall internal losses that is beneficial for application as quantum processing elements or memories. Indeed, high mechanical quality ($Q$) factors have been demonstrated at MHz range frequencies \cite{goryachev2012,Steele2015HighQ,Schiesser2017} in different kinds of devices, with the $Q$ values sometimes exceeding $\sim 10^9$. Moreover, mechanical oscillators can rather straightforwardly be coupled to electromagnetic fields of nearly any reasonable frequency. In particular, they are routinely coupled to propagating fields in optics. The latter coupling takes place via cavity optomechanical interaction, which has been used to study various types of mechanical systems in the quantum limit of motion, for example using membrane or beam oscillators \cite{Teufel2011b,AspelmeyerCool11,Painter2016SiN} coupled to microwave cavities, or coupling optical frequencies to phononic crystal oscillators \cite{AspelmeyerCool11}. One motivation to study optomechanical systems and their hybrids with other quantum devices is the potential of frequency conversion near the quantum limit between microwaves and optics, where promising steps have already been taken \cite{LehnertInterf2014,ClelandInterf,Polzik2014Amp,Srinivasan2016}. 

The types of oscillators mentioned above can in principle be coupled to superconducting qubits, and the coupling has been experimentally demonstrated for beams and membranes \cite{LaHaye2009,transmonnems,Simmonds2015qb,LSETNEMSexp,Santos2017}. There is also an extensive theoretical literature discussing what happens when qubits and mechanics are put together, see e.g.~\cite{Armour2002,Zoller2004,TianPRB2005,ClelandPRA05,Nori2006CPB,Blencowe2007CPB,buksQND,Milburn2009CPB,Fazio2011}. A promising approach to realize the interaction is to use piezoelectric materials that strongly enhance the electromechanical coupling. One can therefore use GHz frequency mechanical modes  that otherwise would have too small interaction energy in order to realize resonant interaction with the qubit. A thin film bulk acoustic wave resonator (FBAR) was strongly coupled to the qubit \cite{ClelandMartinis}, although with diminished $Q$ value. Recently, several experiments have successfully integrated qubits with surface acoustic wave resonances in piezoelectric substrates \cite{Delsing2014,Leek2017SAW,Nakamura2017SAW,Astafiev2017SAW,Lehnert2017SAW}, showing much higher $Q$ values. An alternative option, discussed in the current work, is offered by high-overtone bulk acoustic wave resonators (HBAR) where the mechanical energy is diluted in a low-loss substrate \cite{BAW1979,HBAR1992,HBAR1993,HBAR2006,HBAR2016,Tang2016}. These systems have been investigated because they show promise as a clock source due to their high $Q$ values up to $\sim 10^5$ at GHz frequencies at ambient conditions. A HBAR resonator was recently successfully demonstrated in a 3D transmon architecture \cite{SchoelkopfHBAR2017}. The high quality factor ensures that the acoustic energy stays long in the system giving access to long-living phonon states, whereas piezoelectricity provides a mechanism to transduce displacement into an electrical signal, thus providing an interface between the mechanical modes and the quantum state of a superconducting qubit. In our work we take an alternative approach to realize a HBAR resonator in the basic on-chip design where both the qubit and a measurement cavity are fabricated on-chip. The maximal mode overlap in our design allows for qubit-mechanics coupling of $\sim 5$ MHz, an order of magnitude higher than in Ref.~\cite{SchoelkopfHBAR2017}.

\begin{figure*}[t]
  \begin{center}
    {\includegraphics[width=0.8\textwidth]{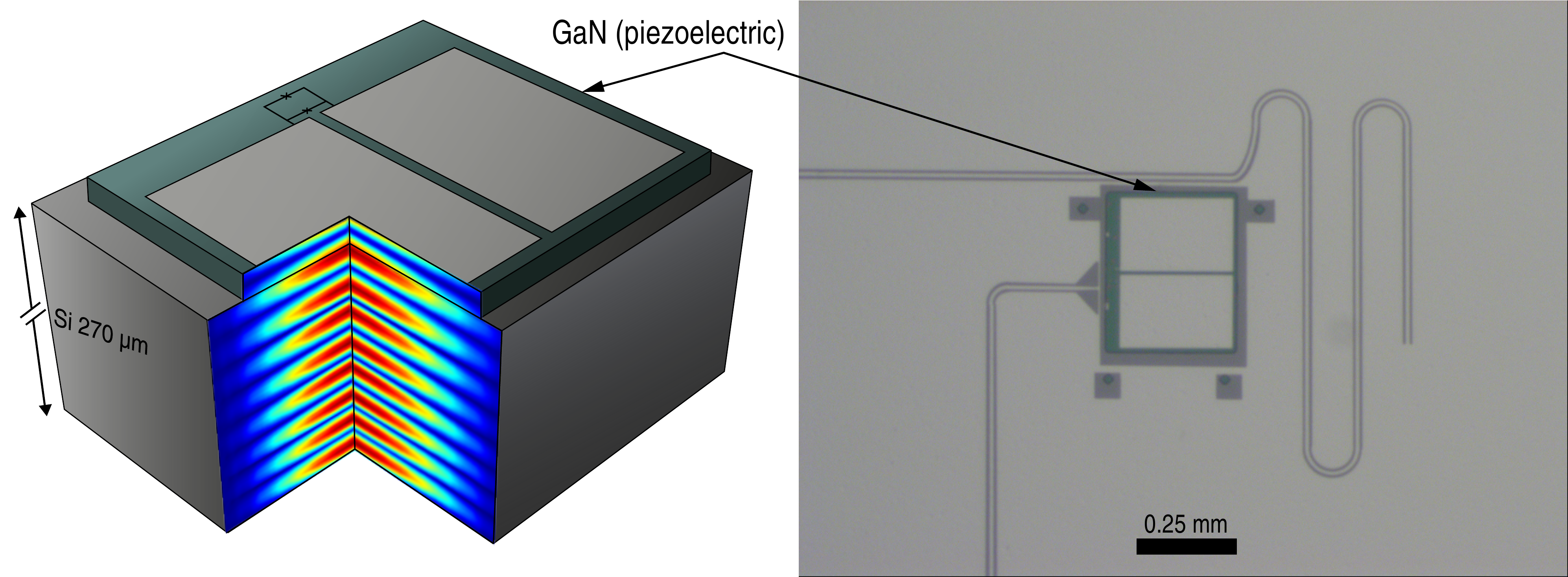}} 
    \caption{\emph{Device schematics.} Left (not to scale): View of the transmon qubit fabricated on top of the patterned GaN piezoelectric layer. The cutout shows the displacement profile for one high-overtone bulk acoustic resonance (HBAR) mode underneath the electrode (in the measurement, the device is operated around a much higher mode with the number of nodes $\gtrsim 300$). Right: corresponding optical micrograph of the measured device. The meandering readout cavity is coupled to the qubit with a coupling energy $\simeq 35$ MHz. The Josephson energy of the qubit can be controlled by the on-chip flux bias line.}
    \label{fig:schematic}
 \end{center}
\end{figure*} 

\section{Device design}

The HBAR resonator, as shown in \fref{fig:schematic} consists of a piezoelectric thin film grown on an acoustically low-loss substrate. Electrodes define acoustic modes that couple to electric fields via the piezoelectric effect.  Excitation of the thin piezoelectric film on top of a low loss substrate creates acoustic longitudinal waves, or thickness modes, that propagates through the body of the whole material stack. The two air-solid interfaces of the substrate surfaces form parallel and reflecting surfaces for energy trapping. The HBAR resonates in thickness mode expanding and contracting in vertical direction generating standing wave resonances confining the energy of the resonance into the structure similar to an optical Fabry-Pérot cavity. Since the mode volume extends throughout the substrate, the system exhibits a dense spectrum of resonances that can be accessed via the electrodes. The piezoelectric layer modulates the amplitudes throughout the frequency spectrum so that the most strongly excited resonances occur when the half wavelength of the acoustic wave is equal to the thickness of the piezoelectric film $t_p$ according to
$f_{0} = v_p/2t_p$  where $v_p$ is the acoustic velocity in the piezoelectric material. Thus, the center frequency of the modes is determined by the thickness of the piezoelectric film whereas the spacing between two consecutive resonances is inversely proportional to the thickness of the substrate. The mode spacing or the free spectral range (FSR) of the resonator is given by
$f_{\m{FSR}} = v_s/2t$, where $v_s$ is the acoustic velocity in the substrate and $t$ is the substrate thickness. Typical substrate thickness is a few hundreds of micrometer so that the frequency spacing is few tens of megahertz. Since the thickness of piezoelectric layer is in the order of a micrometer, HBAR naturally operate in the gigahertz regime that is typical for superconducting qubits or other quantum devices. 

In our device, the  piezoelectric thin film material is gallium nitride (GaN). It is a a wide band-gap semiconductor material that has become more and more popular over the recent years. The unique properties of GaN make it an important material in optoelectronics, and high-power, high speed electronics. However, GaN based micro-electromechanical systems are still largely unexplored \cite{GaNBAR2006,GaNBAR2009,GaNnems2010,GaNreview}. GaN has high acoustic velocities, low elastic losses and a relatively strong piezoelectric effect. In terms of the magnitude of the relevant piezoelectric stress coefficient $e_{33}$, GaN ($e_{33} \sim 0.7$ C/m$^2$) \cite{GaNreview} is in between the common materials AlN ($\sim 1.5$ C/m$^2$) and quartz ($\sim 0.1$ C/m$^2$). Thin films of GaN can be grown epitaxially, and hence one can expect they would have small elastic losses due to a minimal amount of atomic two-level systems abundant in amorphous materials.

The sample that is used in this work has been fabricated from double side polished 2" GaN/Si wafers that have been obtained from Kyma Technologies. The wafers have a layer structure of 500 nm GaN grown epitaxially by hydride vapour phase epitaxy (HVPE), and a small buffer layer of AlN on high-resistivity fz-silicon with $\langle 111 \rangle $ orientation. The fabrication process begins by masking the area for GaN transducer with a SU-8 5 resist mask in preparation for argon ion etching of GaN. The Ar-ion bombardment is used to physically etch away the GaN. The etching is continued until the silicon substrate is reached. The SU-8 is removed with a combination of plasma etch and Piranha solution. The top circuit is defined using e-beam lithography on a bilayer resist. The electrical circuit is formed in a single metallization step using double-angle evaporation.



Photograph and a schematic of the device is displayed in Figure \ref{fig:schematic}. The system consists of patterned GaN piezoelectric transducer, a split Josephson junction transmon qubit, and a microwave coplanar waveguide (CPW) cavity. The transmon qubit is fabricated directly on top of the GaN film. Two large pads of the transmon form the parallel capacitance of the qubit and serve as the electrodes to excite the mechanical resonator. The two electrodes form separate resonating modes under each electrode. However, due to their close proximity, the two resonators are acoustically interacting with each other creating a laterally coupled resonator, so that the resonating system is essentially behaving as a single resonator.

 The transmon is read-out with a quarterwave coplanar waveguide cavity. The CPW cavity has a 
 resonant frequency of $\omega_c/2\pi$ = 6.113  GHz, and an external linewidth of $\gamma_c$ = 2 MHz. The transmon's transition energy can be tuned with an on-chip flux line. The sample is measured in  sample holder that allows free movement of the top and bottom surfaces of the device. The sample holder is enclosed in Cryoperm magnetic shield and attached to the base of a dilution refrigerator with base temperature of 20 mK.

\section{Results}
\begin{figure}[h]
  \begin{center}
    {\includegraphics[width=0.45\textwidth]{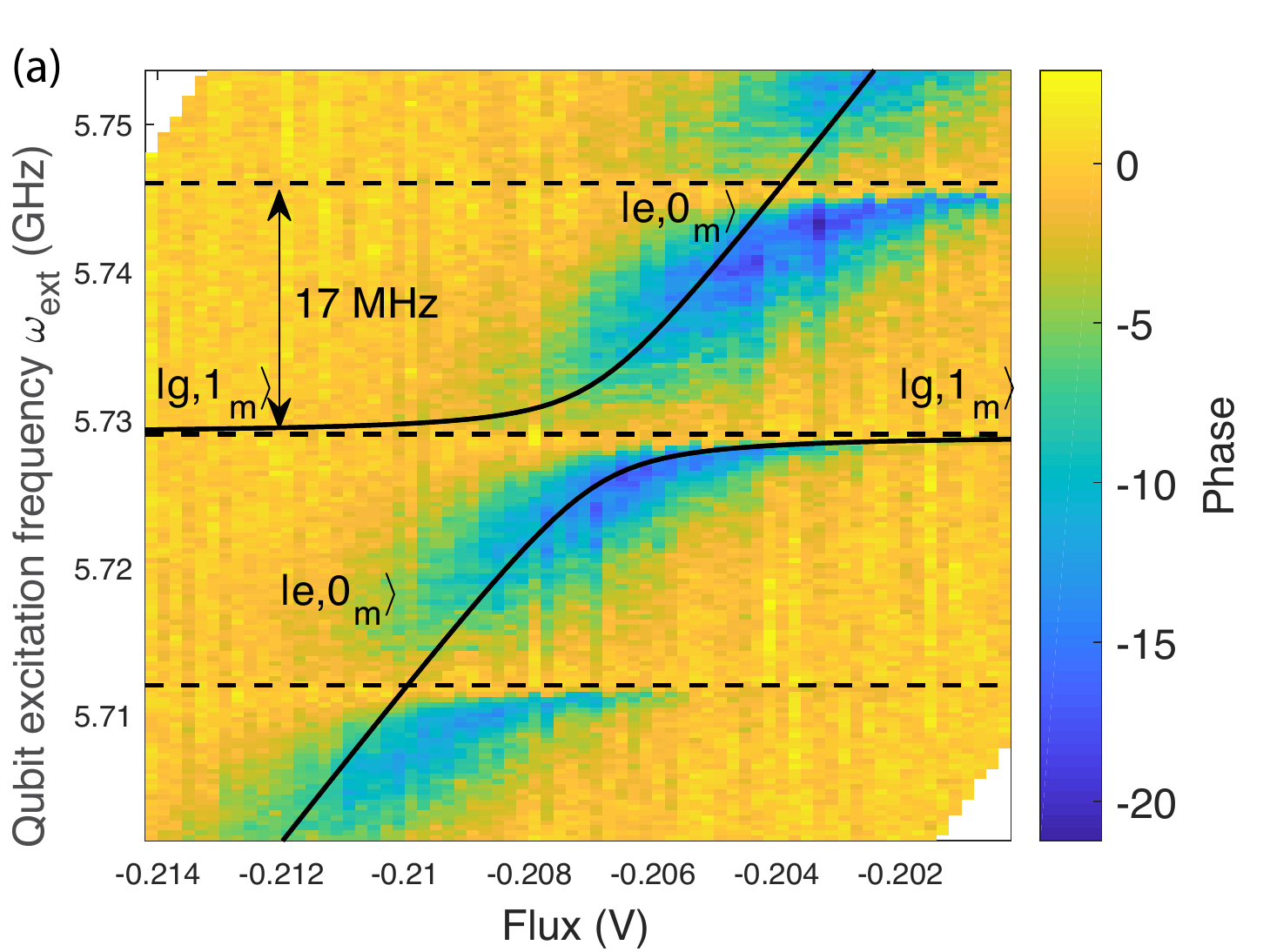}} 
    {\includegraphics[width=0.45\textwidth]{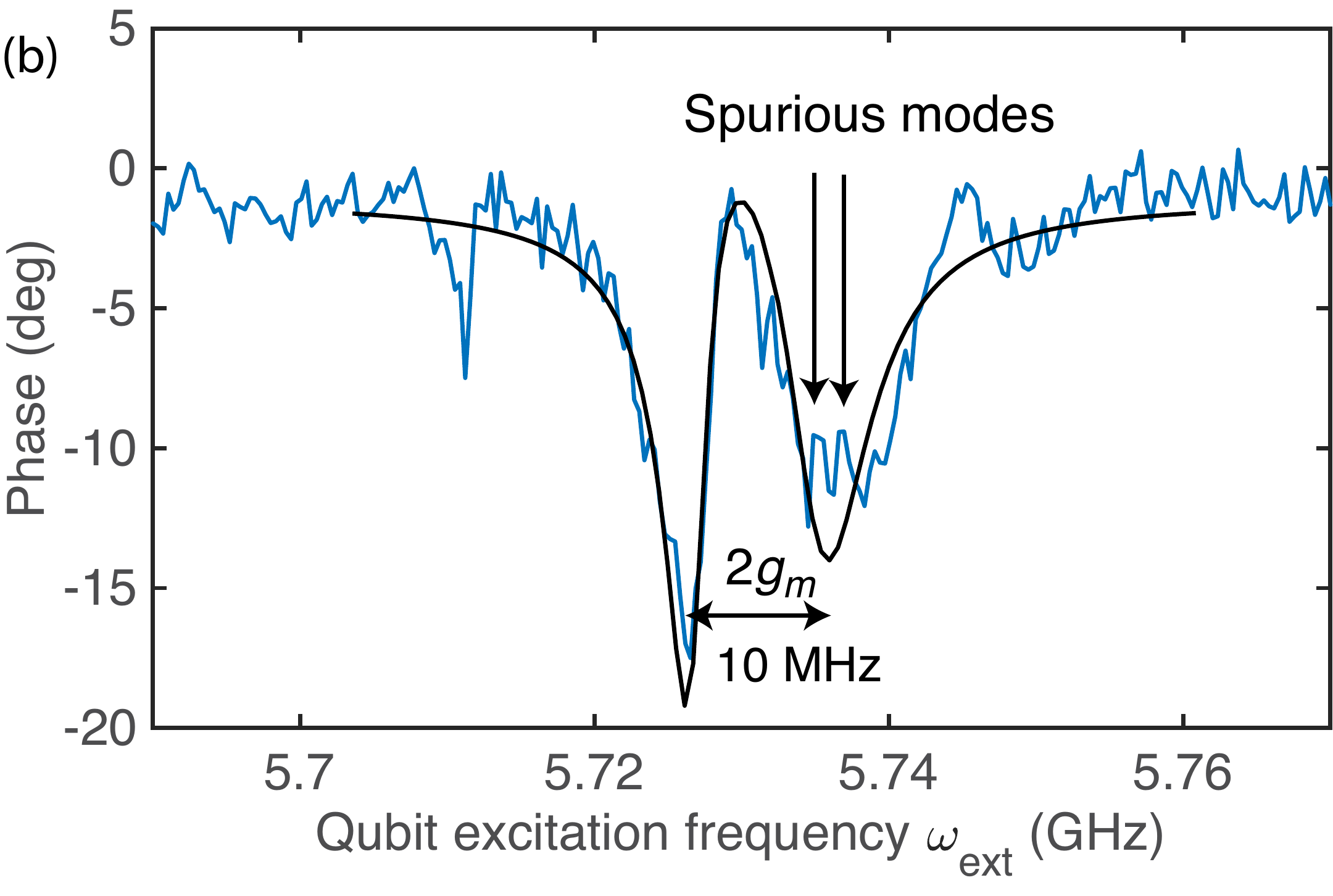}} 
    \caption{\emph{Spectroscopy of qubit-mechanics interaction.} (a) Cavity phase shift in two-tone spectroscopy depicting the qubit population as a function of the qubit excitation frequency and flux bias. The bare qubit frequency scans the panel diagonally from lower left to upper right. The dashed lines indicate overtone HBAR mechanical modes that each couple to the qubit. The expected energy levels for coupling of the qubit to one of the modes ($m = 336$) are drawn on top of the data. (b) Cross-section at the flux bias $V \simeq -0.206$ V showing the qubit vacuum Rabi splitting at the onset of strong coupling $g_m \sim \kappa$. The solid curve is a theoretical fit.}
    \label{fig:coupling}
 \end{center}
\end{figure}

The qubit-phonon coupling was investigated in a standard two-tone spectroscopy measurement, where a drive signal of variable frequency $\omega_{\m{ext}}$ is applied to the flux line. The qubit's transition frequency can be determined from the dispersive shift of the measurement tone. As shown in \fref{fig:coupling}(a), as the qubit frequency is tuned via external magnetic flux, the qubit experiences evenly spaced anticrossing features every 17 MHz. Each anticrossing is due to a resonance of an overtone mechanical mode. The spacing corresponds to a longitudinal sound velocity of $v_s = 9200$ m/s for a measured substrate thickness of 270 $\upmu$m. This agrees for the longitudinal velocity in silicon.

In \fref{fig:coupling}(b), we show a cross-section through one of the resonances around 5.73 GHz, which equals to a longitudinal mode number of 336 and roughly corresponds to the maximum qubit-mechanics coupling where half a wavelength fits in the piezo layer thickness. A fit using our theoretical model (see below) involving coupling of the qubit to one acoustic model, provides a good agreement to the acoustic vacuum Rabi splitting. Based on the fit, the coupling strength between the qubit and an individual acoustic mode can be determined to be $g_m \approx 5$ MHz, which is comparable to the linewidth $\kappa \simeq 8$ MHz of the qubit. This puts the device close to the strong coupling regime, where quantum control of the coupled system becomes possible. The limiting factor for entering more deeply into the strong coupling regime is the enhanced decoherence of the qubit fabricated on the piezoelectric platform. From the theoretical model, we obtain the qubit energy decay and dephasing times $T_1 \simeq 20$ ns, and $T_2 \simeq 400$ ns, respectively.


Adjacent to the each acoustic mode in \fref{fig:coupling}(a) is a number of faint, overlapping peaks spaced of the order 1 MHz. These are the acoustic spurious modes that arise due to geometry of the electrodes, and they appear on the higher frequency side of the main acoustic mode. The spurious modes have complicated mode profiles, and they each couple only weakly to the qubit. However, a large number of spurious modes act to broaden the linewidth of the qubit on the higher-frequency side of each anticrossing. We take the spurious modes phenomenologically into account in the simulation shown in  \fref{fig:coupling}(b) by setting the qubit decay rate to be enhanced by 20 \% to the right of the resonance, a value used as an adjustable parameter in the analysis.

\begin{figure}[h]
  \begin{center}
    {\includegraphics[width=0.45\textwidth]{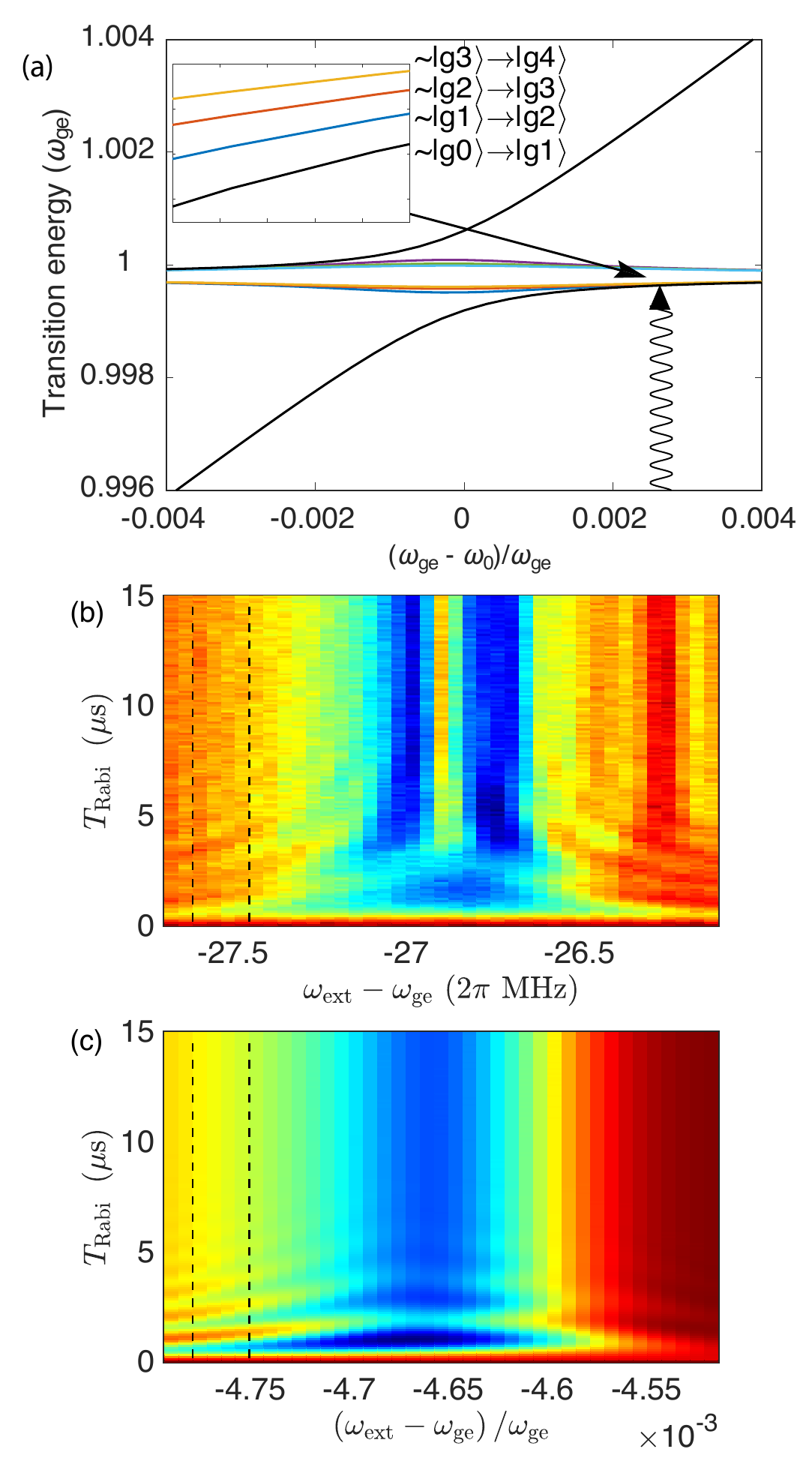}} 
    \caption{\emph{Rabi oscillations.} (a) Energy levels in the near-resonant tail of the Jaynes-Cummings spectrum used for demonstrating coherent oscillations in the electromechanical system. The inset shows zoom-in of the nearly harmonic transitions that have only a small qubit component. (b) Measured Rabi oscillations of the qubit population as a function of excitation frequency. The drive amplitude is $\Omega/2\pi \simeq 6$ MHz. The dashed vertical lines indicate positions of the line cuts plotted in \fref{fig:fock}. (c) Corresponding numerical simulation.}
    \label{fig:rabis}
 \end{center}
\end{figure}

Next, we investigated the electromechanical system in the time domain measurements, allowing to verify the possibility for a coherent control. Long lifetime of the main acoustic modes show promise for time-domain quantum operations. However,  since the qubit total decoherence is strongly enhanced due to coupling of a multitude of spurious acoustic modes, we explore the qubit-mechanics interaction in an uncommon parameter regime that allows for long-lasting coherent oscillations in spite of fast qubit decay. Namely, we introduce a significant detuning $\Delta = \omega_{\m{ge}} - \omega_{m}$ between the qubit (frequency $\omega_{\m{ge}}$) and an acoustic mode in question (frequency $\omega_{m}$), by the amount of several $g_m$'s (see \fref{fig:rabis}(a)). In this limit, the Jaynes-Cummings energy eigenstates 
\begin{equation}\label{eq:JCvect}
\begin{split}
|+,n \rangle =& \cos \left( \Theta_n/2 \right) |e,n
\rangle + \sin \left( \Theta_n/2 \right) | g, n+1 \rangle \\
|-,n \rangle =& - \sin \left( \Theta_n/2 \right) |e,n
\rangle + \cos\left( \Theta_n/2 \right) | g, n+1 \rangle
\end{split}
\end{equation}
represent only marginally mixed qubit and oscillator states. Here, $\tan \Theta_n = 2 g_m \sqrt{n+1}/\Delta$, and $g$ and $e$ are the qubit ground and excited states, and $n$ denotes the oscillator phonon number. 

In the detuned limit, the oscillator-type transitions occur between the states of approximate form $|-,n \rangle \simeq | g, n+1 \rangle$, and the energies $E_{-,n} =  \omega_m (n+1) - \frac{1}{2}\sqrt{4g_m^2(n+1)+\Delta^2}$. The qubit relaxation matrix element is suppressed down to $(n+1)(g_m/\Delta)^2$. Even in this detuned state, some anharmonicity is preserved, allowing for driving coherent oscillations that flop energy between the driven qubit and several acoustic Fock states of the oscillator. The oscillator state, however,  resembles a driven linear oscillator, and individual access to the Fock states is limited.



As illustrated in \fref{fig:rabis}(a), we perform a Rabi oscillation measurement by applying microwave excitation to the flux line near-resonant to the oscillator-like transitions. The system is first excited with a drive pulse and the probability of the excited state is monitored as a function of Rabi pulse width $T_{\m{Rabi}}$. The results of the excited state population in the Rabi measurement are shown in Figure \ref{fig:rabis}(b). 
Two distinct peaks in the center in the figure can be identified as the modes under each electrode. Nanometer size differences in the piezoelectric layer or electrode height can result in such a lifting of the degeneracy.

The results are analyzed by simulating a qubit-oscillator interaction by numerically solving the Liouvillean master equation for a qubit coupled to one of the acoustic harmonic modes (index $m$). We operate in the rotating frame defined by the qubit excitation frequency. The Hamiltonian is
\begin{equation}
H = -\dfrac{1}{2} \Delta_{\m{ext}} \sigma_z + \omega_m a_m^\dagger a_m + g_m \left(a_m \sigma_{+} + a_m^\dagger\sigma_{-} \right) + \frac{\Omega}{2}  \sigma_x \,,
\end{equation}
where $a_m^\dagger$ and $a_m$ are the creation and annihilation operators for the phonons, whereas $\sigma_z$, $\sigma_{+}$ and $\sigma_{-}$ represent the qubit operators. $\Delta_{\m{ext}} = \omega_{\m{ge}} - \omega_{\m{ext}}$ is the drive detuning, and $\Omega$ is the Rabi frequency of the  drive. The oscillator is initially thought to be well in its ground state since $k_bT/\hbar \omega_m \ll 1$. We incorporate standard Lindblad operators for the qubit including decay and dephasing, and for the oscillator with a decay rate $\simeq 60$ kHz.



\begin{figure}[h]
  \begin{center}
    {\includegraphics[width=0.45\textwidth]{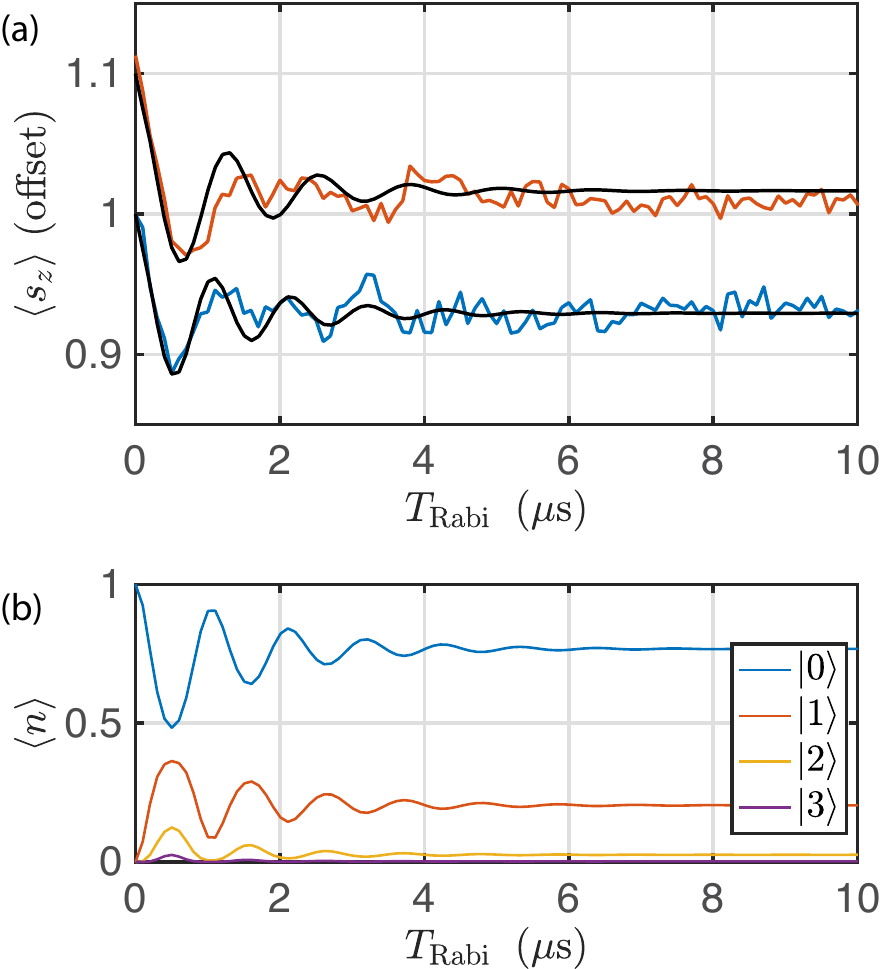}} 
    \caption{\emph{Coherent oscillations.} (a) Qubit excited state population at drive frequencies $\Delta_{\m{ext}}/2\pi \simeq 27.7$ MHz (lower) and $\Delta_{\m{ext}}/2\pi \simeq 27.5$ MHz (upper). The upper curve is vertically offset by 0.1 units for clarity. The black solid curves are from the theoretical model. (b) Theoretically predicted phonon Fock state occupations corresponding to (a).}
    \label{fig:fock}
 \end{center}
\end{figure} 

Since the model includes only one harmonic oscillator mode, the spurious peaks seen in \fref{fig:rabis}(b), or influence of the spurious modes on the coherent oscillations cannot be reproduced by the simulation. Otherwise, as shown in \fref{fig:rabis}(c) and \fref{fig:fock}(a), the simulation presents a reasonable agreement to the measurement. We attribute the small discrepancies visible in \fref{fig:fock}(a) to the other oscillator modes. In \fref{fig:fock}(b), we display the expected Fock state occupations corresponding to \fref{fig:fock}(a). We observe the Fock states oscillate in pace with the qubit population, however, given states cannot be constructed by state transfer from the qubit at high fidelity.


\section{Conclusions and future directions}

We have demonstrated here a potential platform for resonantly coupling superconducting qubits to harmonic mechanical modes. The high-overtone bulk acoustic wave mechanical resonator is ideal for interfacing to superconducting quantum devices due to its simple fabrication process and high quality factor even at GHz frequencies. The device that has been studied in this work is affected by  decoherence due to enhanced energy decay. The dissipation can be likely attributed to the spurious acoustic modes that provide additional loss channels thus increasing the dissipations in the qubit. A common technique for attenuating the spurious modes is to have no parallel edges on the piezoelectric transducer, a technique called apodization. Having no parallel edges on the transducer can lead to an improvement in resonator's performance by suppressing the strength of the lateral modes. With non-parallel sides, a laterally traveling acoustic wave leaving from one point on a edge will not get reflected back onto the same spot by the opposite side. Thus the resonant path becomes greater and the lateral standing waves become more attenuated. 
By improving the device the system can be brought deeply into the strong coupling regime. With stronger coupling, mapping the qubit states into a dense memory formed by the various mechanical overtone modes becomes possible, as well as interfacing to other systems via the mechanical modes. One can also consider the mechanical modes as macroscopic quantum systems, and hence one can use advanced quantum protocols to synthesize arbitrary quantum state \cite{Martinis2009} in the mechanical resonator.

\begin{acknowledgments} This work was supported by the Academy of Finland (contract 250280, CoE LTQ, 275245), by the European Research Council (615755-CAVITYQPD), and by the Centre for Quantum Engineering at Aalto University. We acknowledge funding from the European Union’s Horizon 2020 research and innovation program under grant agreement No. 732894 (FETPRO HOT). The work benefited from the facilities at the OtaNano - Micronova Nanofabrication Center and at the Low Temperature Laboratory.
\end{acknowledgments}

\bibliography{hbar}

\begin{thebibliography}{46}%
\makeatletter
\providecommand \@ifxundefined [1]{%
 \@ifx{#1\undefined}
}%
\providecommand \@ifnum [1]{%
 \ifnum #1\expandafter \@firstoftwo
 \else \expandafter \@secondoftwo
 \fi
}%
\providecommand \@ifx [1]{%
 \ifx #1\expandafter \@firstoftwo
 \else \expandafter \@secondoftwo
 \fi
}%
\providecommand \natexlab [1]{#1}%
\providecommand \enquote  [1]{``#1''}%
\providecommand \bibnamefont  [1]{#1}%
\providecommand \bibfnamefont [1]{#1}%
\providecommand \citenamefont [1]{#1}%
\providecommand \href@noop [0]{\@secondoftwo}%
\providecommand \href [0]{\begingroup \@sanitize@url \@href}%
\providecommand \@href[1]{\@@startlink{#1}\@@href}%
\providecommand \@@href[1]{\endgroup#1\@@endlink}%
\providecommand \@sanitize@url [0]{\catcode `\\12\catcode `\$12\catcode
  `\&12\catcode `\#12\catcode `\^12\catcode `\_12\catcode `\%12\relax}%
\providecommand \@@startlink[1]{}%
\providecommand \@@endlink[0]{}%
\providecommand \url  [0]{\begingroup\@sanitize@url \@url }%
\providecommand \@url [1]{\endgroup\@href {#1}{\urlprefix }}%
\providecommand \urlprefix  [0]{URL }%
\providecommand \Eprint [0]{\href }%
\providecommand \doibase [0]{http://dx.doi.org/}%
\providecommand \selectlanguage [0]{\@gobble}%
\providecommand \bibinfo  [0]{\@secondoftwo}%
\providecommand \bibfield  [0]{\@secondoftwo}%
\providecommand \translation [1]{[#1]}%
\providecommand \BibitemOpen [0]{}%
\providecommand \bibitemStop [0]{}%
\providecommand \bibitemNoStop [0]{.\EOS\space}%
\providecommand \EOS [0]{\spacefactor3000\relax}%
\providecommand \BibitemShut  [1]{\csname bibitem#1\endcsname}%
\let\auto@bib@innerbib\@empty
\bibitem [{\citenamefont {Wendin}(2017)}]{WendinReview2017}%
  \BibitemOpen
  \bibfield  {author} {\bibinfo {author} {\bibfnamefont {G}~\bibnamefont
  {Wendin}},\ }\bibfield  {title} {\enquote {\bibinfo {title} {Quantum
  information processing with superconducting circuits: a review},}\
  }\href@noop {} {\bibfield  {journal} {\bibinfo  {journal} {Reports on
  Progress in Physics}\ }\textbf {\bibinfo {volume} {80}},\ \bibinfo {pages}
  {106001} (\bibinfo {year} {2017})}\BibitemShut {NoStop}%
\bibitem [{\citenamefont {Zhu}\ \emph {et~al.}(2011)\citenamefont {Zhu},
  \citenamefont {Saito}, \citenamefont {Kemp}, \citenamefont {Kakuyanagi},
  \citenamefont {Karimoto}, \citenamefont {Nakano}, \citenamefont {Munro},
  \citenamefont {Tokura}, \citenamefont {Everitt}, \citenamefont {Nemoto},
  \citenamefont {Kasu}, \citenamefont {Mizuochi},\ and\ \citenamefont
  {Semba}}]{SembaQBspin}%
  \BibitemOpen
  \bibfield  {author} {\bibinfo {author} {\bibfnamefont {Xiaobo}\ \bibnamefont
  {Zhu}}, \bibinfo {author} {\bibfnamefont {Shiro}\ \bibnamefont {Saito}},
  \bibinfo {author} {\bibfnamefont {Alexander}\ \bibnamefont {Kemp}}, \bibinfo
  {author} {\bibfnamefont {Kosuke}\ \bibnamefont {Kakuyanagi}}, \bibinfo
  {author} {\bibfnamefont {Shin-ichi}\ \bibnamefont {Karimoto}}, \bibinfo
  {author} {\bibfnamefont {Hayato}\ \bibnamefont {Nakano}}, \bibinfo {author}
  {\bibfnamefont {William~J.}\ \bibnamefont {Munro}}, \bibinfo {author}
  {\bibfnamefont {Yasuhiro}\ \bibnamefont {Tokura}}, \bibinfo {author}
  {\bibfnamefont {Mark~S.}\ \bibnamefont {Everitt}}, \bibinfo {author}
  {\bibfnamefont {Kae}\ \bibnamefont {Nemoto}}, \bibinfo {author}
  {\bibfnamefont {Makoto}\ \bibnamefont {Kasu}}, \bibinfo {author}
  {\bibfnamefont {Norikazu}\ \bibnamefont {Mizuochi}}, \ and\ \bibinfo {author}
  {\bibfnamefont {Kouichi}\ \bibnamefont {Semba}},\ }\bibfield  {title}
  {\enquote {\bibinfo {title} {Coherent coupling of a superconducting flux
  qubit to an electron spin ensemble in diamond},}\ }\href@noop {} {\bibfield
  {journal} {\bibinfo  {journal} {Nature}\ }\textbf {\bibinfo {volume} {478}},\
  \bibinfo {pages} {221--224} (\bibinfo {year} {2011})}\BibitemShut {NoStop}%
\bibitem [{\citenamefont {Kubo}\ \emph {et~al.}(2011)\citenamefont {Kubo},
  \citenamefont {Grezes}, \citenamefont {Dewes}, \citenamefont {Umeda},
  \citenamefont {Isoya}, \citenamefont {Sumiya}, \citenamefont {Morishita},
  \citenamefont {Abe}, \citenamefont {Onoda}, \citenamefont {Ohshima},
  \citenamefont {Jacques}, \citenamefont {Dr\'eau}, \citenamefont {Roch},
  \citenamefont {Diniz}, \citenamefont {Auffeves}, \citenamefont {Vion},
  \citenamefont {Esteve},\ and\ \citenamefont {Bertet}}]{EsteveSpin2011}%
  \BibitemOpen
  \bibfield  {author} {\bibinfo {author} {\bibfnamefont {Y.}~\bibnamefont
  {Kubo}}, \bibinfo {author} {\bibfnamefont {C.}~\bibnamefont {Grezes}},
  \bibinfo {author} {\bibfnamefont {A.}~\bibnamefont {Dewes}}, \bibinfo
  {author} {\bibfnamefont {T.}~\bibnamefont {Umeda}}, \bibinfo {author}
  {\bibfnamefont {J.}~\bibnamefont {Isoya}}, \bibinfo {author} {\bibfnamefont
  {H.}~\bibnamefont {Sumiya}}, \bibinfo {author} {\bibfnamefont
  {N.}~\bibnamefont {Morishita}}, \bibinfo {author} {\bibfnamefont
  {H.}~\bibnamefont {Abe}}, \bibinfo {author} {\bibfnamefont {S.}~\bibnamefont
  {Onoda}}, \bibinfo {author} {\bibfnamefont {T.}~\bibnamefont {Ohshima}},
  \bibinfo {author} {\bibfnamefont {V.}~\bibnamefont {Jacques}}, \bibinfo
  {author} {\bibfnamefont {A.}~\bibnamefont {Dr\'eau}}, \bibinfo {author}
  {\bibfnamefont {J.-F.}\ \bibnamefont {Roch}}, \bibinfo {author}
  {\bibfnamefont {I.}~\bibnamefont {Diniz}}, \bibinfo {author} {\bibfnamefont
  {A.}~\bibnamefont {Auffeves}}, \bibinfo {author} {\bibfnamefont
  {D.}~\bibnamefont {Vion}}, \bibinfo {author} {\bibfnamefont {D.}~\bibnamefont
  {Esteve}}, \ and\ \bibinfo {author} {\bibfnamefont {P.}~\bibnamefont
  {Bertet}},\ }\bibfield  {title} {\enquote {\bibinfo {title} {Hybrid quantum
  circuit with a superconducting qubit coupled to a spin ensemble},}\
  }\href@noop {} {\bibfield  {journal} {\bibinfo  {journal} {Phys. Rev. Lett.}\
  }\textbf {\bibinfo {volume} {107}},\ \bibinfo {pages} {220501} (\bibinfo
  {year} {2011})}\BibitemShut {NoStop}%
\bibitem [{\citenamefont {Tabuchi}\ \emph {et~al.}(2015)\citenamefont
  {Tabuchi}, \citenamefont {Ishino}, \citenamefont {Noguchi}, \citenamefont
  {Ishikawa}, \citenamefont {Yamazaki}, \citenamefont {Usami},\ and\
  \citenamefont {Nakamura}}]{Tabuchi2015}%
  \BibitemOpen
  \bibfield  {author} {\bibinfo {author} {\bibfnamefont {Yutaka}\ \bibnamefont
  {Tabuchi}}, \bibinfo {author} {\bibfnamefont {Seiichiro}\ \bibnamefont
  {Ishino}}, \bibinfo {author} {\bibfnamefont {Atsushi}\ \bibnamefont
  {Noguchi}}, \bibinfo {author} {\bibfnamefont {Toyofumi}\ \bibnamefont
  {Ishikawa}}, \bibinfo {author} {\bibfnamefont {Rekishu}\ \bibnamefont
  {Yamazaki}}, \bibinfo {author} {\bibfnamefont {Koji}\ \bibnamefont {Usami}},
  \ and\ \bibinfo {author} {\bibfnamefont {Yasunobu}\ \bibnamefont
  {Nakamura}},\ }\bibfield  {title} {\enquote {\bibinfo {title} {Coherent
  coupling between a ferromagnetic magnon and a superconducting qubit},}\
  }\href@noop {} {\bibfield  {journal} {\bibinfo  {journal} {Science}\ }\textbf
  {\bibinfo {volume} {349}},\ \bibinfo {pages} {405--408} (\bibinfo {year}
  {2015})}\BibitemShut {NoStop}%
\bibitem [{\citenamefont {Goryachev}\ \emph {et~al.}(2012)\citenamefont
  {Goryachev}, \citenamefont {Creedon}, \citenamefont {Ivanov}, \citenamefont
  {Galliou}, \citenamefont {Bourquin},\ and\ \citenamefont
  {Tobar}}]{goryachev2012}%
  \BibitemOpen
  \bibfield  {author} {\bibinfo {author} {\bibfnamefont {Maxim}\ \bibnamefont
  {Goryachev}}, \bibinfo {author} {\bibfnamefont {Daniel~L}\ \bibnamefont
  {Creedon}}, \bibinfo {author} {\bibfnamefont {Eugene~N}\ \bibnamefont
  {Ivanov}}, \bibinfo {author} {\bibfnamefont {Serge}\ \bibnamefont {Galliou}},
  \bibinfo {author} {\bibfnamefont {Roger}\ \bibnamefont {Bourquin}}, \ and\
  \bibinfo {author} {\bibfnamefont {Michael~E}\ \bibnamefont {Tobar}},\
  }\bibfield  {title} {\enquote {\bibinfo {title} {Extremely low-loss acoustic
  phonons in a quartz bulk acoustic wave resonator at millikelvin
  temperature},}\ }\href@noop {} {\bibfield  {journal} {\bibinfo  {journal}
  {Appl.~Phys.~Lett.}\ }\textbf {\bibinfo {volume} {100}},\ \bibinfo {pages}
  {243504} (\bibinfo {year} {2012})}\BibitemShut {NoStop}%
\bibitem [{\citenamefont {Yuan}\ \emph {et~al.}(2015)\citenamefont {Yuan},
  \citenamefont {Cohen},\ and\ \citenamefont {Steele}}]{Steele2015HighQ}%
  \BibitemOpen
  \bibfield  {author} {\bibinfo {author} {\bibfnamefont {Mingyun}\ \bibnamefont
  {Yuan}}, \bibinfo {author} {\bibfnamefont {Martijn~A.}\ \bibnamefont
  {Cohen}}, \ and\ \bibinfo {author} {\bibfnamefont {Gary~A.}\ \bibnamefont
  {Steele}},\ }\bibfield  {title} {\enquote {\bibinfo {title} {Silicon nitride
  membrane resonators at millikelvin temperatures with quality factors
  exceeding $10^8$},}\ }\href@noop {} {\bibfield  {journal} {\bibinfo
  {journal} {Applied Physics Letters}\ }\textbf {\bibinfo {volume} {107}},\
  \bibinfo {pages} {263501} (\bibinfo {year} {2015})}\BibitemShut {NoStop}%
\bibitem [{\citenamefont {Tsaturyan}\ \emph {et~al.}(2017)\citenamefont
  {Tsaturyan}, \citenamefont {Barg}, \citenamefont {Polzik},\ and\
  \citenamefont {Schliesser}}]{Schiesser2017}%
  \BibitemOpen
  \bibfield  {author} {\bibinfo {author} {\bibfnamefont {Y.}~\bibnamefont
  {Tsaturyan}}, \bibinfo {author} {\bibfnamefont {A.}~\bibnamefont {Barg}},
  \bibinfo {author} {\bibfnamefont {E.~S.}\ \bibnamefont {Polzik}}, \ and\
  \bibinfo {author} {\bibfnamefont {A.}~\bibnamefont {Schliesser}},\ }\bibfield
   {title} {\enquote {\bibinfo {title} {Ultracoherent nanomechanical resonators
  via soft clamping and dissipation dilution},}\ }\href@noop {} {\bibfield
  {journal} {\bibinfo  {journal} {Nature Nanotechnology}\ }\textbf {\bibinfo
  {volume} {12}},\ \bibinfo {pages} {776} (\bibinfo {year} {2017})}\BibitemShut
  {NoStop}%
\bibitem [{\citenamefont {Teufel}\ \emph {et~al.}(2011)\citenamefont {Teufel},
  \citenamefont {Donner}, \citenamefont {Li}, \citenamefont {Harlow},
  \citenamefont {Allman}, \citenamefont {Cicak}, \citenamefont {Sirois},
  \citenamefont {Whittaker}, \citenamefont {Lehnert},\ and\ \citenamefont
  {Simmonds}}]{Teufel2011b}%
  \BibitemOpen
  \bibfield  {author} {\bibinfo {author} {\bibfnamefont {J.~D.}\ \bibnamefont
  {Teufel}}, \bibinfo {author} {\bibfnamefont {T.}~\bibnamefont {Donner}},
  \bibinfo {author} {\bibfnamefont {Dale}\ \bibnamefont {Li}}, \bibinfo
  {author} {\bibfnamefont {J.~W.}\ \bibnamefont {Harlow}}, \bibinfo {author}
  {\bibfnamefont {M.~S.}\ \bibnamefont {Allman}}, \bibinfo {author}
  {\bibfnamefont {K.}~\bibnamefont {Cicak}}, \bibinfo {author} {\bibfnamefont
  {A.~J.}\ \bibnamefont {Sirois}}, \bibinfo {author} {\bibfnamefont {J.~D.}\
  \bibnamefont {Whittaker}}, \bibinfo {author} {\bibfnamefont {K.~W.}\
  \bibnamefont {Lehnert}}, \ and\ \bibinfo {author} {\bibfnamefont {R.~W.}\
  \bibnamefont {Simmonds}},\ }\bibfield  {title} {\enquote {\bibinfo {title}
  {{Sideband cooling of micromechanical motion to the quantum ground state}},}\
  }\href@noop {} {\bibfield  {journal} {\bibinfo  {journal} {Nature}\ }\textbf
  {\bibinfo {volume} {475}},\ \bibinfo {pages} {359--363} (\bibinfo {year}
  {2011})}\BibitemShut {NoStop}%
\bibitem [{\citenamefont {Chan}\ \emph {et~al.}(2011)\citenamefont {Chan},
  \citenamefont {Alegre}, \citenamefont {Safavi-Naeini}, \citenamefont {Hill},
  \citenamefont {Krause}, \citenamefont {Gr\"oblacher}, \citenamefont
  {Aspelmeyer},\ and\ \citenamefont {Painter}}]{AspelmeyerCool11}%
  \BibitemOpen
  \bibfield  {author} {\bibinfo {author} {\bibfnamefont {Jasper}\ \bibnamefont
  {Chan}}, \bibinfo {author} {\bibfnamefont {T.~P.~Mayer}\ \bibnamefont
  {Alegre}}, \bibinfo {author} {\bibfnamefont {Amir~H.}\ \bibnamefont
  {Safavi-Naeini}}, \bibinfo {author} {\bibfnamefont {Jeff~T.}\ \bibnamefont
  {Hill}}, \bibinfo {author} {\bibfnamefont {Alex}\ \bibnamefont {Krause}},
  \bibinfo {author} {\bibfnamefont {Simon}\ \bibnamefont {Gr\"oblacher}},
  \bibinfo {author} {\bibfnamefont {Markus}\ \bibnamefont {Aspelmeyer}}, \ and\
  \bibinfo {author} {\bibfnamefont {Oskar}\ \bibnamefont {Painter}},\
  }\bibfield  {title} {\enquote {\bibinfo {title} {Laser cooling of a
  nanomechanical oscillator into its quantum ground state},}\ }\href@noop {}
  {\bibfield  {journal} {\bibinfo  {journal} {Nature}\ }\textbf {\bibinfo
  {volume} {478}},\ \bibinfo {pages} {89--92} (\bibinfo {year}
  {2011})}\BibitemShut {NoStop}%
\bibitem [{\citenamefont {Fink}\ \emph {et~al.}(2016)\citenamefont {Fink},
  \citenamefont {Kalaee}, \citenamefont {Pitanti}, \citenamefont {Norte},
  \citenamefont {Heinzle}, \citenamefont {Davan{\c c}o}, \citenamefont
  {Srinivasan},\ and\ \citenamefont {Painter}}]{Painter2016SiN}%
  \BibitemOpen
  \bibfield  {author} {\bibinfo {author} {\bibfnamefont {J.~M.}\ \bibnamefont
  {Fink}}, \bibinfo {author} {\bibfnamefont {M.}~\bibnamefont {Kalaee}},
  \bibinfo {author} {\bibfnamefont {A.}~\bibnamefont {Pitanti}}, \bibinfo
  {author} {\bibfnamefont {R.}~\bibnamefont {Norte}}, \bibinfo {author}
  {\bibfnamefont {L.}~\bibnamefont {Heinzle}}, \bibinfo {author} {\bibfnamefont
  {M.}~\bibnamefont {Davan{\c c}o}}, \bibinfo {author} {\bibfnamefont
  {K.}~\bibnamefont {Srinivasan}}, \ and\ \bibinfo {author} {\bibfnamefont
  {O.}~\bibnamefont {Painter}},\ }\bibfield  {title} {\enquote {\bibinfo
  {title} {Quantum electromechanics on silicon nitride nanomembranes},}\
  }\href@noop {} {\bibfield  {journal} {\bibinfo  {journal} {Nature
  Communications}\ }\textbf {\bibinfo {volume} {7}},\ \bibinfo {pages} {12396}
  (\bibinfo {year} {2016})}\BibitemShut {NoStop}%
\bibitem [{\citenamefont {Andrews}\ \emph {et~al.}(2014)\citenamefont
  {Andrews}, \citenamefont {Peterson}, \citenamefont {Purdy}, \citenamefont
  {Cicak}, \citenamefont {Simmonds}, \citenamefont {Regal},\ and\ \citenamefont
  {Lehnert}}]{LehnertInterf2014}%
  \BibitemOpen
  \bibfield  {author} {\bibinfo {author} {\bibfnamefont {R.~W.}\ \bibnamefont
  {Andrews}}, \bibinfo {author} {\bibfnamefont {R.~W.}\ \bibnamefont
  {Peterson}}, \bibinfo {author} {\bibfnamefont {T.~P.}\ \bibnamefont {Purdy}},
  \bibinfo {author} {\bibfnamefont {K.}~\bibnamefont {Cicak}}, \bibinfo
  {author} {\bibfnamefont {R.~W.}\ \bibnamefont {Simmonds}}, \bibinfo {author}
  {\bibfnamefont {C.~A.}\ \bibnamefont {Regal}}, \ and\ \bibinfo {author}
  {\bibfnamefont {K.~W.}\ \bibnamefont {Lehnert}},\ }\bibfield  {title}
  {\enquote {\bibinfo {title} {Bidirectional and efficient conversion between
  microwave and optical light},}\ }\href@noop {} {\bibfield  {journal}
  {\bibinfo  {journal} {Nature Physics}\ }\textbf {\bibinfo {volume} {10}},\
  \bibinfo {pages} {321} (\bibinfo {year} {2014})}\BibitemShut {NoStop}%
\bibitem [{\citenamefont {Bochmann}\ \emph {et~al.}(2013)\citenamefont
  {Bochmann}, \citenamefont {Vainsencher}, \citenamefont {Awschalom},\ and\
  \citenamefont {Cleland}}]{ClelandInterf}%
  \BibitemOpen
  \bibfield  {author} {\bibinfo {author} {\bibfnamefont {Joerg}\ \bibnamefont
  {Bochmann}}, \bibinfo {author} {\bibfnamefont {Amit}\ \bibnamefont
  {Vainsencher}}, \bibinfo {author} {\bibfnamefont {David~D.}\ \bibnamefont
  {Awschalom}}, \ and\ \bibinfo {author} {\bibfnamefont {Andrew~N.}\
  \bibnamefont {Cleland}},\ }\bibfield  {title} {\enquote {\bibinfo {title}
  {Nanomechanical coupling between microwave and optical photons},}\
  }\href@noop {} {\bibfield  {journal} {\bibinfo  {journal} {Nat. Phys.}\
  }\textbf {\bibinfo {volume} {9}},\ \bibinfo {pages} {712--716} (\bibinfo
  {year} {2013})}\BibitemShut {NoStop}%
\bibitem [{\citenamefont {Bagci}\ \emph {et~al.}(2014)\citenamefont {Bagci},
  \citenamefont {Simonsen}, \citenamefont {Schmid}, \citenamefont {Villanueva},
  \citenamefont {Zeuthen}, \citenamefont {Appel}, \citenamefont {Taylor},
  \citenamefont {Sorensen}, \citenamefont {Usami}, \citenamefont {Schliesser},\
  and\ \citenamefont {Polzik}}]{Polzik2014Amp}%
  \BibitemOpen
  \bibfield  {author} {\bibinfo {author} {\bibfnamefont {T.}~\bibnamefont
  {Bagci}}, \bibinfo {author} {\bibfnamefont {A.}~\bibnamefont {Simonsen}},
  \bibinfo {author} {\bibfnamefont {S.}~\bibnamefont {Schmid}}, \bibinfo
  {author} {\bibfnamefont {L.~G.}\ \bibnamefont {Villanueva}}, \bibinfo
  {author} {\bibfnamefont {E.}~\bibnamefont {Zeuthen}}, \bibinfo {author}
  {\bibfnamefont {J.}~\bibnamefont {Appel}}, \bibinfo {author} {\bibfnamefont
  {J.~M.}\ \bibnamefont {Taylor}}, \bibinfo {author} {\bibfnamefont
  {A.}~\bibnamefont {Sorensen}}, \bibinfo {author} {\bibfnamefont
  {K.}~\bibnamefont {Usami}}, \bibinfo {author} {\bibfnamefont
  {A.}~\bibnamefont {Schliesser}}, \ and\ \bibinfo {author} {\bibfnamefont
  {E.~S.}\ \bibnamefont {Polzik}},\ }\bibfield  {title} {\enquote {\bibinfo
  {title} {Optical detection of radio waves through a nanomechanical
  transducer},}\ }\href@noop {} {\bibfield  {journal} {\bibinfo  {journal}
  {Nature}\ }\textbf {\bibinfo {volume} {507}},\ \bibinfo {pages} {81--85}
  (\bibinfo {year} {2014})}\BibitemShut {NoStop}%
\bibitem [{\citenamefont {Balram}\ \emph {et~al.}(2016)\citenamefont {Balram},
  \citenamefont {Davan{\c c}o}, \citenamefont {Song},\ and\ \citenamefont
  {Srinivasan}}]{Srinivasan2016}%
  \BibitemOpen
  \bibfield  {author} {\bibinfo {author} {\bibfnamefont {Krishna~C.}\
  \bibnamefont {Balram}}, \bibinfo {author} {\bibfnamefont {Marcelo~I.}\
  \bibnamefont {Davan{\c c}o}}, \bibinfo {author} {\bibfnamefont {Jin~Dong}\
  \bibnamefont {Song}}, \ and\ \bibinfo {author} {\bibfnamefont {Kartik}\
  \bibnamefont {Srinivasan}},\ }\bibfield  {title} {\enquote {\bibinfo {title}
  {Coherent coupling between radiofrequency, optical and acoustic waves in
  piezo-optomechanical circuits},}\ }\href@noop {} {\bibfield  {journal}
  {\bibinfo  {journal} {Nature Photonics}\ }\textbf {\bibinfo {volume} {10}},\
  \bibinfo {pages} {346} (\bibinfo {year} {2016})}\BibitemShut {NoStop}%
\bibitem [{\citenamefont {LaHaye}\ \emph {et~al.}(2009)\citenamefont {LaHaye},
  \citenamefont {Suh}, \citenamefont {Echternach}, \citenamefont {Schwab},\
  and\ \citenamefont {Roukes}}]{LaHaye2009}%
  \BibitemOpen
  \bibfield  {author} {\bibinfo {author} {\bibfnamefont {M.~D.}\ \bibnamefont
  {LaHaye}}, \bibinfo {author} {\bibfnamefont {J.}~\bibnamefont {Suh}},
  \bibinfo {author} {\bibfnamefont {P.~M.}\ \bibnamefont {Echternach}},
  \bibinfo {author} {\bibfnamefont {K.~C.}\ \bibnamefont {Schwab}}, \ and\
  \bibinfo {author} {\bibfnamefont {M.~L.}\ \bibnamefont {Roukes}},\ }\bibfield
   {title} {\enquote {\bibinfo {title} {Nanomechanical measurements of a
  superconducting qubit},}\ }\href@noop {} {\bibfield  {journal} {\bibinfo
  {journal} {Nature}\ }\textbf {\bibinfo {volume} {459}},\ \bibinfo {pages}
  {960--964} (\bibinfo {year} {2009})}\BibitemShut {NoStop}%
\bibitem [{\citenamefont {Pirkkalainen}\ \emph {et~al.}(2013)\citenamefont
  {Pirkkalainen}, \citenamefont {Cho}, \citenamefont {Li}, \citenamefont
  {Paraoanu}, \citenamefont {Hakonen},\ and\ \citenamefont
  {Sillanp{\"a}{\"a}}}]{transmonnems}%
  \BibitemOpen
  \bibfield  {author} {\bibinfo {author} {\bibfnamefont {J.~M.}\ \bibnamefont
  {Pirkkalainen}}, \bibinfo {author} {\bibfnamefont {S.~U.}\ \bibnamefont
  {Cho}}, \bibinfo {author} {\bibfnamefont {Jian}\ \bibnamefont {Li}}, \bibinfo
  {author} {\bibfnamefont {G.~S.}\ \bibnamefont {Paraoanu}}, \bibinfo {author}
  {\bibfnamefont {P.~J.}\ \bibnamefont {Hakonen}}, \ and\ \bibinfo {author}
  {\bibfnamefont {M.~A.}\ \bibnamefont {Sillanp{\"a}{\"a}}},\ }\bibfield
  {title} {\enquote {\bibinfo {title} {Hybrid circuit cavity quantum
  electrodynamics with a micromechanical resonator},}\ }\href@noop {}
  {\bibfield  {journal} {\bibinfo  {journal} {Nature}\ }\textbf {\bibinfo
  {volume} {494}},\ \bibinfo {pages} {211--215} (\bibinfo {year}
  {2013})}\BibitemShut {NoStop}%
\bibitem [{\citenamefont {Lecocq}\ \emph {et~al.}(2015)\citenamefont {Lecocq},
  \citenamefont {Teufel}, \citenamefont {Aumentado},\ and\ \citenamefont
  {Simmonds}}]{Simmonds2015qb}%
  \BibitemOpen
  \bibfield  {author} {\bibinfo {author} {\bibfnamefont {F.}~\bibnamefont
  {Lecocq}}, \bibinfo {author} {\bibfnamefont {J.~D.}\ \bibnamefont {Teufel}},
  \bibinfo {author} {\bibfnamefont {J.}~\bibnamefont {Aumentado}}, \ and\
  \bibinfo {author} {\bibfnamefont {R.~W.}\ \bibnamefont {Simmonds}},\
  }\bibfield  {title} {\enquote {\bibinfo {title} {Resolving the vacuum
  fluctuations of an optomechanical system using an artificial atom},}\
  }\href@noop {} {\bibfield  {journal} {\bibinfo  {journal} {Nat. Phys.}\
  }\textbf {\bibinfo {volume} {11}},\ \bibinfo {pages} {635--639} (\bibinfo
  {year} {2015})}\BibitemShut {NoStop}%
\bibitem [{\citenamefont {Pirkkalainen}\ \emph {et~al.}(2015)\citenamefont
  {Pirkkalainen}, \citenamefont {Cho}, \citenamefont {Massel}, \citenamefont
  {Tuorila}, \citenamefont {Heikkila}, \citenamefont {Hakonen},\ and\
  \citenamefont {Sillanp\"a\"a}}]{LSETNEMSexp}%
  \BibitemOpen
  \bibfield  {author} {\bibinfo {author} {\bibfnamefont {J.~M.}\ \bibnamefont
  {Pirkkalainen}}, \bibinfo {author} {\bibfnamefont {S.~U.}\ \bibnamefont
  {Cho}}, \bibinfo {author} {\bibfnamefont {F.}~\bibnamefont {Massel}},
  \bibinfo {author} {\bibfnamefont {J.}~\bibnamefont {Tuorila}}, \bibinfo
  {author} {\bibfnamefont {T.~T.}\ \bibnamefont {Heikkila}}, \bibinfo {author}
  {\bibfnamefont {P.~J.}\ \bibnamefont {Hakonen}}, \ and\ \bibinfo {author}
  {\bibfnamefont {M.~A.}\ \bibnamefont {Sillanp\"a\"a}},\ }\bibfield  {title}
  {\enquote {\bibinfo {title} {Cavity optomechanics mediated by a quantum
  two-level system},}\ }\href@noop {} {\bibfield  {journal} {\bibinfo
  {journal} {Nature Communications}\ }\textbf {\bibinfo {volume} {6}},\
  \bibinfo {pages} {6981} (\bibinfo {year} {2015})}\BibitemShut {NoStop}%
\bibitem [{\citenamefont {Santos}\ \emph {et~al.}(2017)\citenamefont {Santos},
  \citenamefont {Li}, \citenamefont {Ilves}, \citenamefont {Ockeloen-Korppi},\
  and\ \citenamefont {Sillanp\"a\"a}}]{Santos2017}%
  \BibitemOpen
  \bibfield  {author} {\bibinfo {author} {\bibfnamefont {J~T}\ \bibnamefont
  {Santos}}, \bibinfo {author} {\bibfnamefont {J}~\bibnamefont {Li}}, \bibinfo
  {author} {\bibfnamefont {J}~\bibnamefont {Ilves}}, \bibinfo {author}
  {\bibfnamefont {C~F}\ \bibnamefont {Ockeloen-Korppi}}, \ and\ \bibinfo
  {author} {\bibfnamefont {M.}~\bibnamefont {Sillanp\"a\"a}},\ }\bibfield
  {title} {\enquote {\bibinfo {title} {Optomechanical measurement of a
  millimeter-sized mechanical oscillator approaching the quantum ground
  state},}\ }\href@noop {} {\bibfield  {journal} {\bibinfo  {journal} {New
  Journal of Physics}\ }\textbf {\bibinfo {volume} {19}},\ \bibinfo {pages}
  {103014} (\bibinfo {year} {2017})}\BibitemShut {NoStop}%
\bibitem [{\citenamefont {Armour}\ \emph {et~al.}(2002)\citenamefont {Armour},
  \citenamefont {Blencowe},\ and\ \citenamefont {Schwab}}]{Armour2002}%
  \BibitemOpen
  \bibfield  {author} {\bibinfo {author} {\bibfnamefont {A.~D.}\ \bibnamefont
  {Armour}}, \bibinfo {author} {\bibfnamefont {M.~P.}\ \bibnamefont
  {Blencowe}}, \ and\ \bibinfo {author} {\bibfnamefont {K.~C.}\ \bibnamefont
  {Schwab}},\ }\bibfield  {title} {\enquote {\bibinfo {title} {Entanglement and
  decoherence of a micromechanical resonator via coupling to a cooper-pair
  box},}\ }\href@noop {} {\bibfield  {journal} {\bibinfo  {journal} {Phys. Rev.
  Lett.}\ }\textbf {\bibinfo {volume} {88}},\ \bibinfo {pages} {148301}
  (\bibinfo {year} {2002})}\BibitemShut {NoStop}%
\bibitem [{\citenamefont {Rabl}\ \emph {et~al.}(2004)\citenamefont {Rabl},
  \citenamefont {Shnirman},\ and\ \citenamefont {Zoller}}]{Zoller2004}%
  \BibitemOpen
  \bibfield  {author} {\bibinfo {author} {\bibfnamefont {P.}~\bibnamefont
  {Rabl}}, \bibinfo {author} {\bibfnamefont {A.}~\bibnamefont {Shnirman}}, \
  and\ \bibinfo {author} {\bibfnamefont {P.}~\bibnamefont {Zoller}},\
  }\bibfield  {title} {\enquote {\bibinfo {title} {Generation of squeezed
  states of nanomechanical resonators by reservoir engineering},}\ }\href@noop
  {} {\bibfield  {journal} {\bibinfo  {journal} {Phys. Rev. B}\ }\textbf
  {\bibinfo {volume} {70}},\ \bibinfo {pages} {205304} (\bibinfo {year}
  {2004})}\BibitemShut {NoStop}%
\bibitem [{\citenamefont {Tian}(2005)}]{TianPRB2005}%
  \BibitemOpen
  \bibfield  {author} {\bibinfo {author} {\bibfnamefont {L.}~\bibnamefont
  {Tian}},\ }\bibfield  {title} {\enquote {\bibinfo {title} {{Entanglement from
  a nanomechanical resonator weakly coupled to a single Cooper-pair box}},}\
  }\href@noop {} {\bibfield  {journal} {\bibinfo  {journal} {Phys. Rev. B}\
  }\textbf {\bibinfo {volume} {72}},\ \bibinfo {pages} {195411} (\bibinfo
  {year} {2005})}\BibitemShut {NoStop}%
\bibitem [{\citenamefont {Geller}\ and\ \citenamefont
  {Cleland}(2005)}]{ClelandPRA05}%
  \BibitemOpen
  \bibfield  {author} {\bibinfo {author} {\bibfnamefont {M.~R.}\ \bibnamefont
  {Geller}}\ and\ \bibinfo {author} {\bibfnamefont {A.~N.}\ \bibnamefont
  {Cleland}},\ }\bibfield  {title} {\enquote {\bibinfo {title} {Superconducting
  qubits coupled to nanoelectromechanical resonators: An architecture for
  solid-state quantum-information processing},}\ }\href@noop {} {\bibfield
  {journal} {\bibinfo  {journal} {Phys. Rev. A}\ }\textbf {\bibinfo {volume}
  {71}},\ \bibinfo {pages} {032311} (\bibinfo {year} {2005})}\BibitemShut
  {NoStop}%
\bibitem [{\citenamefont {Wei}\ \emph {et~al.}(2006)\citenamefont {Wei},
  \citenamefont {Liu}, \citenamefont {Sun},\ and\ \citenamefont
  {Nori}}]{Nori2006CPB}%
  \BibitemOpen
  \bibfield  {author} {\bibinfo {author} {\bibfnamefont {L.~F.}\ \bibnamefont
  {Wei}}, \bibinfo {author} {\bibfnamefont {Yu-xi}\ \bibnamefont {Liu}},
  \bibinfo {author} {\bibfnamefont {C.~P.}\ \bibnamefont {Sun}}, \ and\
  \bibinfo {author} {\bibfnamefont {Franco}\ \bibnamefont {Nori}},\ }\bibfield
  {title} {\enquote {\bibinfo {title} {Probing tiny motions of nanomechanical
  resonators: Classical or quantum mechanical?}}\ }\href@noop {} {\bibfield
  {journal} {\bibinfo  {journal} {Phys. Rev. Lett.}\ }\textbf {\bibinfo
  {volume} {97}},\ \bibinfo {pages} {237201} (\bibinfo {year}
  {2006})}\BibitemShut {NoStop}%
\bibitem [{\citenamefont {Jacobs}\ \emph {et~al.}(2007)\citenamefont {Jacobs},
  \citenamefont {Lougovski},\ and\ \citenamefont {Blencowe}}]{Blencowe2007CPB}%
  \BibitemOpen
  \bibfield  {author} {\bibinfo {author} {\bibfnamefont {Kurt}\ \bibnamefont
  {Jacobs}}, \bibinfo {author} {\bibfnamefont {Pavel}\ \bibnamefont
  {Lougovski}}, \ and\ \bibinfo {author} {\bibfnamefont {Miles}\ \bibnamefont
  {Blencowe}},\ }\bibfield  {title} {\enquote {\bibinfo {title} {Continuous
  measurement of the energy eigenstates of a nanomechanical resonator without a
  nondemolition probe},}\ }\href@noop {} {\bibfield  {journal} {\bibinfo
  {journal} {Phys. Rev. Lett.}\ }\textbf {\bibinfo {volume} {98}},\ \bibinfo
  {pages} {147201} (\bibinfo {year} {2007})}\BibitemShut {NoStop}%
\bibitem [{\citenamefont {Buks}\ \emph {et~al.}(2008)\citenamefont {Buks},
  \citenamefont {Segev}, \citenamefont {Zaitsev}, \citenamefont {Abdo},\ and\
  \citenamefont {Blencowe}}]{buksQND}%
  \BibitemOpen
  \bibfield  {author} {\bibinfo {author} {\bibfnamefont {E.}~\bibnamefont
  {Buks}}, \bibinfo {author} {\bibfnamefont {E.}~\bibnamefont {Segev}},
  \bibinfo {author} {\bibfnamefont {S.}~\bibnamefont {Zaitsev}}, \bibinfo
  {author} {\bibfnamefont {B.}~\bibnamefont {Abdo}}, \ and\ \bibinfo {author}
  {\bibfnamefont {M.~P.}\ \bibnamefont {Blencowe}},\ }\bibfield  {title}
  {\enquote {\bibinfo {title} {Quantum nondemolition measurement of discrete
  fock states of a nanomechanical resonator},}\ }\href@noop {} {\bibfield
  {journal} {\bibinfo  {journal} {Europhysics Letters}\ }\textbf {\bibinfo
  {volume} {81}},\ \bibinfo {pages} {10001} (\bibinfo {year}
  {2008})}\BibitemShut {NoStop}%
\bibitem [{\citenamefont {Semi\~ao}\ \emph {et~al.}(2009)\citenamefont
  {Semi\~ao}, \citenamefont {Furuya},\ and\ \citenamefont
  {Milburn}}]{Milburn2009CPB}%
  \BibitemOpen
  \bibfield  {author} {\bibinfo {author} {\bibfnamefont {F.~L.}\ \bibnamefont
  {Semi\~ao}}, \bibinfo {author} {\bibfnamefont {K.}~\bibnamefont {Furuya}}, \
  and\ \bibinfo {author} {\bibfnamefont {G.~J.}\ \bibnamefont {Milburn}},\
  }\bibfield  {title} {\enquote {\bibinfo {title} {Kerr nonlinearities and
  nonclassical states with superconducting qubits and nanomechanical
  resonators},}\ }\href@noop {} {\bibfield  {journal} {\bibinfo  {journal}
  {Phys. Rev. A}\ }\textbf {\bibinfo {volume} {79}},\ \bibinfo {pages} {063811}
  (\bibinfo {year} {2009})}\BibitemShut {NoStop}%
\bibitem [{\citenamefont {Didier}\ \emph {et~al.}(2011)\citenamefont {Didier},
  \citenamefont {Pugnetti}, \citenamefont {Blanter},\ and\ \citenamefont
  {Fazio}}]{Fazio2011}%
  \BibitemOpen
  \bibfield  {author} {\bibinfo {author} {\bibfnamefont {Nicolas}\ \bibnamefont
  {Didier}}, \bibinfo {author} {\bibfnamefont {Stefano}\ \bibnamefont
  {Pugnetti}}, \bibinfo {author} {\bibfnamefont {Yaroslav~M.}\ \bibnamefont
  {Blanter}}, \ and\ \bibinfo {author} {\bibfnamefont {Rosario}\ \bibnamefont
  {Fazio}},\ }\bibfield  {title} {\enquote {\bibinfo {title} {Detecting phonon
  blockade with photons},}\ }\href@noop {} {\bibfield  {journal} {\bibinfo
  {journal} {Phys. Rev. B}\ }\textbf {\bibinfo {volume} {84}},\ \bibinfo
  {pages} {054503} (\bibinfo {year} {2011})}\BibitemShut {NoStop}%
\bibitem [{\citenamefont {O'Connell}\ \emph {et~al.}(2010)\citenamefont
  {O'Connell}, \citenamefont {Hofheinz}, \citenamefont {Ansmann}, \citenamefont
  {Bialczak}, \citenamefont {Lenander}, \citenamefont {Lucero}, \citenamefont
  {Neeley}, \citenamefont {Sank}, \citenamefont {Wang}, \citenamefont {Weides},
  \citenamefont {Wenner}, \citenamefont {Martinis},\ and\ \citenamefont
  {Cleland}}]{ClelandMartinis}%
  \BibitemOpen
  \bibfield  {author} {\bibinfo {author} {\bibfnamefont {A~D}\ \bibnamefont
  {O'Connell}}, \bibinfo {author} {\bibfnamefont {M}~\bibnamefont {Hofheinz}},
  \bibinfo {author} {\bibfnamefont {M}~\bibnamefont {Ansmann}}, \bibinfo
  {author} {\bibfnamefont {Radoslaw~C}\ \bibnamefont {Bialczak}}, \bibinfo
  {author} {\bibfnamefont {M}~\bibnamefont {Lenander}}, \bibinfo {author}
  {\bibfnamefont {Erik}\ \bibnamefont {Lucero}}, \bibinfo {author}
  {\bibfnamefont {M}~\bibnamefont {Neeley}}, \bibinfo {author} {\bibfnamefont
  {D}~\bibnamefont {Sank}}, \bibinfo {author} {\bibfnamefont {H}~\bibnamefont
  {Wang}}, \bibinfo {author} {\bibfnamefont {M}~\bibnamefont {Weides}},
  \bibinfo {author} {\bibfnamefont {J}~\bibnamefont {Wenner}}, \bibinfo
  {author} {\bibfnamefont {John~M}\ \bibnamefont {Martinis}}, \ and\ \bibinfo
  {author} {\bibfnamefont {A~N}\ \bibnamefont {Cleland}},\ }\bibfield  {title}
  {\enquote {\bibinfo {title} {{Quantum ground state and single-phonon control
  of a mechanical resonator}},}\ }\href@noop {} {\bibfield  {journal} {\bibinfo
   {journal} {Nature}\ }\textbf {\bibinfo {volume} {464}},\ \bibinfo {pages}
  {697--703} (\bibinfo {year} {2010})}\BibitemShut {NoStop}%
\bibitem [{\citenamefont {Gustafsson}\ \emph {et~al.}(2014)\citenamefont
  {Gustafsson}, \citenamefont {Aref}, \citenamefont {Kockum}, \citenamefont
  {Ekstr\"om}, \citenamefont {Johansson},\ and\ \citenamefont
  {Delsing}}]{Delsing2014}%
  \BibitemOpen
  \bibfield  {author} {\bibinfo {author} {\bibfnamefont {Martin~V.}\
  \bibnamefont {Gustafsson}}, \bibinfo {author} {\bibfnamefont {Thomas}\
  \bibnamefont {Aref}}, \bibinfo {author} {\bibfnamefont {Anton~Frisk}\
  \bibnamefont {Kockum}}, \bibinfo {author} {\bibfnamefont {Maria~K.}\
  \bibnamefont {Ekstr\"om}}, \bibinfo {author} {\bibfnamefont {Göran}\
  \bibnamefont {Johansson}}, \ and\ \bibinfo {author} {\bibfnamefont {Per}\
  \bibnamefont {Delsing}},\ }\bibfield  {title} {\enquote {\bibinfo {title}
  {Propagating phonons coupled to an artificial atom},}\ }\href@noop {}
  {\bibfield  {journal} {\bibinfo  {journal} {Science}\ }\textbf {\bibinfo
  {volume} {346}},\ \bibinfo {pages} {207--211} (\bibinfo {year}
  {2014})}\BibitemShut {NoStop}%
\bibitem [{\citenamefont {Manenti}\ \emph {et~al.}(2017)\citenamefont
  {Manenti}, \citenamefont {Kockum}, \citenamefont {Patterson}, \citenamefont
  {Behrle}, \citenamefont {Rahamim}, \citenamefont {Tancredi}, \citenamefont
  {Nori},\ and\ \citenamefont {Leek}}]{Leek2017SAW}%
  \BibitemOpen
  \bibfield  {author} {\bibinfo {author} {\bibfnamefont {Riccardo}\
  \bibnamefont {Manenti}}, \bibinfo {author} {\bibfnamefont {Anton~F.}\
  \bibnamefont {Kockum}}, \bibinfo {author} {\bibfnamefont {Andrew}\
  \bibnamefont {Patterson}}, \bibinfo {author} {\bibfnamefont {Tanja}\
  \bibnamefont {Behrle}}, \bibinfo {author} {\bibfnamefont {Joseph}\
  \bibnamefont {Rahamim}}, \bibinfo {author} {\bibfnamefont {Giovanna}\
  \bibnamefont {Tancredi}}, \bibinfo {author} {\bibfnamefont {Franco}\
  \bibnamefont {Nori}}, \ and\ \bibinfo {author} {\bibfnamefont {Peter~J.}\
  \bibnamefont {Leek}},\ }\bibfield  {title} {\enquote {\bibinfo {title}
  {Circuit quantum acoustodynamics with surface acoustic waves},}\ }\href@noop
  {} {\bibfield  {journal} {\bibinfo  {journal} {Nature Communications}\
  }\textbf {\bibinfo {volume} {8}},\ \bibinfo {pages} {975} (\bibinfo {year}
  {2017})}\BibitemShut {NoStop}%
\bibitem [{\citenamefont {Noguchi}\ \emph {et~al.}(2017)\citenamefont
  {Noguchi}, \citenamefont {Yamazaki}, \citenamefont {Tabuchi},\ and\
  \citenamefont {Nakamura}}]{Nakamura2017SAW}%
  \BibitemOpen
  \bibfield  {author} {\bibinfo {author} {\bibfnamefont {Atsushi}\ \bibnamefont
  {Noguchi}}, \bibinfo {author} {\bibfnamefont {Rekishu}\ \bibnamefont
  {Yamazaki}}, \bibinfo {author} {\bibfnamefont {Yutaka}\ \bibnamefont
  {Tabuchi}}, \ and\ \bibinfo {author} {\bibfnamefont {Yasunobu}\ \bibnamefont
  {Nakamura}},\ }\bibfield  {title} {\enquote {\bibinfo {title} {Qubit-assisted
  transduction for a detection of surface acoustic waves near the quantum
  limit},}\ }\href@noop {} {\bibfield  {journal} {\bibinfo  {journal} {Phys.
  Rev. Lett.}\ }\textbf {\bibinfo {volume} {119}},\ \bibinfo {pages} {180505}
  (\bibinfo {year} {2017})}\BibitemShut {NoStop}%
\bibitem [{\citenamefont {Bolgar}\ \emph {et~al.}(2017)\citenamefont {Bolgar},
  \citenamefont {Zotova}, \citenamefont {Kirichenko}, \citenamefont {Besedin},
  \citenamefont {Semenov}, \citenamefont {Shaikhaidarov},\ and\ \citenamefont
  {Astafiev}}]{Astafiev2017SAW}%
  \BibitemOpen
  \bibfield  {author} {\bibinfo {author} {\bibfnamefont {A.~N.}\ \bibnamefont
  {Bolgar}}, \bibinfo {author} {\bibfnamefont {J.~I.}\ \bibnamefont {Zotova}},
  \bibinfo {author} {\bibfnamefont {D.~D.}\ \bibnamefont {Kirichenko}},
  \bibinfo {author} {\bibfnamefont {I.~S.}\ \bibnamefont {Besedin}}, \bibinfo
  {author} {\bibfnamefont {A.~V.}\ \bibnamefont {Semenov}}, \bibinfo {author}
  {\bibfnamefont {R.~S.}\ \bibnamefont {Shaikhaidarov}}, \ and\ \bibinfo
  {author} {\bibfnamefont {O.~V.}\ \bibnamefont {Astafiev}},\ }\bibfield
  {title} {\enquote {\bibinfo {title} {Experimental demonstration of a
  two-dimensional phonon cavity in the quantum regime},}\ }\href@noop {}
  {\bibfield  {journal} {\bibinfo  {journal} {arXiv:1710.06476}\ } (\bibinfo
  {year} {2017})}\BibitemShut {NoStop}%
\bibitem [{\citenamefont {Moores}\ \emph {et~al.}(2017)\citenamefont {Moores},
  \citenamefont {Sletten}, \citenamefont {Viennot},\ and\ \citenamefont
  {Lehnert}}]{Lehnert2017SAW}%
  \BibitemOpen
  \bibfield  {author} {\bibinfo {author} {\bibfnamefont {Bradley~A.}\
  \bibnamefont {Moores}}, \bibinfo {author} {\bibfnamefont {Lucas~R.}\
  \bibnamefont {Sletten}}, \bibinfo {author} {\bibfnamefont {Jeremie~J.}\
  \bibnamefont {Viennot}}, \ and\ \bibinfo {author} {\bibfnamefont {K.~W.}\
  \bibnamefont {Lehnert}},\ }\bibfield  {title} {\enquote {\bibinfo {title}
  {Cavity quantum acoustic device in the multimode strong coupling regime},}\
  }\href@noop {} {\bibfield  {journal} {\bibinfo  {journal} {arXiv:1711.05913}\
  } (\bibinfo {year} {2017})}\BibitemShut {NoStop}%
\bibitem [{\citenamefont {Moore}\ \emph {et~al.}(1979)\citenamefont {Moore},
  \citenamefont {Hopwood}, \citenamefont {Haynes},\ and\ \citenamefont
  {McAvoy}}]{BAW1979}%
  \BibitemOpen
  \bibfield  {author} {\bibinfo {author} {\bibfnamefont {R.~A.}\ \bibnamefont
  {Moore}}, \bibinfo {author} {\bibfnamefont {F.~W.}\ \bibnamefont {Hopwood}},
  \bibinfo {author} {\bibfnamefont {T.}~\bibnamefont {Haynes}}, \ and\ \bibinfo
  {author} {\bibfnamefont {B.~R.}\ \bibnamefont {McAvoy}},\ }\bibfield  {title}
  {\enquote {\bibinfo {title} {Bulk acoustic resonators for microwave
  frequencies},}\ }in\ \href@noop {} {\emph {\bibinfo {booktitle} {33rd Annual
  Symposium on Frequency Control}}}\ (\bibinfo {year} {1979})\ pp.\ \bibinfo
  {pages} {444--448}\BibitemShut {NoStop}%
\bibitem [{\citenamefont {Bailey}\ \emph {et~al.}(1992)\citenamefont {Bailey},
  \citenamefont {Driscoll}, \citenamefont {Jelen},\ and\ \citenamefont
  {McAvoy}}]{HBAR1992}%
  \BibitemOpen
  \bibfield  {author} {\bibinfo {author} {\bibfnamefont {D.~S.}\ \bibnamefont
  {Bailey}}, \bibinfo {author} {\bibfnamefont {M.~M.}\ \bibnamefont
  {Driscoll}}, \bibinfo {author} {\bibfnamefont {R.~A.}\ \bibnamefont {Jelen}},
  \ and\ \bibinfo {author} {\bibfnamefont {B.~R.}\ \bibnamefont {McAvoy}},\
  }\bibfield  {title} {\enquote {\bibinfo {title} {Frequency stability of
  high-overtone bulk-acoustic resonators},}\ }\href@noop {} {\bibfield
  {journal} {\bibinfo  {journal} {IEEE Transactions on Ultrasonics,
  Ferroelectrics, and Frequency Control}\ }\textbf {\bibinfo {volume} {39}},\
  \bibinfo {pages} {780--784} (\bibinfo {year} {1992})}\BibitemShut {NoStop}%
\bibitem [{\citenamefont {Kline}\ \emph {et~al.}(1993)\citenamefont {Kline},
  \citenamefont {Lakin},\ and\ \citenamefont {McCarron}}]{HBAR1993}%
  \BibitemOpen
  \bibfield  {author} {\bibinfo {author} {\bibfnamefont {G.~R.}\ \bibnamefont
  {Kline}}, \bibinfo {author} {\bibfnamefont {K.~M.}\ \bibnamefont {Lakin}}, \
  and\ \bibinfo {author} {\bibfnamefont {K.~T.}\ \bibnamefont {McCarron}},\
  }\bibfield  {title} {\enquote {\bibinfo {title} {{Overmoded high Q resonators
  for microwave oscillators}},}\ }in\ \href@noop {} {\emph {\bibinfo
  {booktitle} {1993 IEEE International Frequency Control Symposium}}}\
  (\bibinfo {year} {1993})\ pp.\ \bibinfo {pages} {718--721}\BibitemShut
  {NoStop}%
\bibitem [{\citenamefont {Zhang}\ \emph {et~al.}(2006)\citenamefont {Zhang},
  \citenamefont {Pang}, \citenamefont {Yu},\ and\ \citenamefont
  {Kim}}]{HBAR2006}%
  \BibitemOpen
  \bibfield  {author} {\bibinfo {author} {\bibfnamefont {Hao}\ \bibnamefont
  {Zhang}}, \bibinfo {author} {\bibfnamefont {Wei}\ \bibnamefont {Pang}},
  \bibinfo {author} {\bibfnamefont {Hongyu}\ \bibnamefont {Yu}}, \ and\
  \bibinfo {author} {\bibfnamefont {Eun~Sok}\ \bibnamefont {Kim}},\ }\bibfield
  {title} {\enquote {\bibinfo {title} {High-tone bulk acoustic resonators on
  sapphire, crystal quartz, fused silica, and silicon substrates},}\
  }\href@noop {} {\bibfield  {journal} {\bibinfo  {journal} {Journal of Applied
  Physics}\ }\textbf {\bibinfo {volume} {99}},\ \bibinfo {pages} {124911}
  (\bibinfo {year} {2006})}\BibitemShut {NoStop}%
\bibitem [{\citenamefont {Li}\ \emph {et~al.}(2016)\citenamefont {Li},
  \citenamefont {Liu},\ and\ \citenamefont {Wang}}]{HBAR2016}%
  \BibitemOpen
  \bibfield  {author} {\bibinfo {author} {\bibfnamefont {Jian}\ \bibnamefont
  {Li}}, \bibinfo {author} {\bibfnamefont {Mengwei}\ \bibnamefont {Liu}}, \
  and\ \bibinfo {author} {\bibfnamefont {Chenghao}\ \bibnamefont {Wang}},\
  }\bibfield  {title} {\enquote {\bibinfo {title} {Resonance spectrum
  characteristics of effective electromechanical coupling coefficient of
  high-overtone bulk acoustic resonator},}\ }\href@noop {} {\bibfield
  {journal} {\bibinfo  {journal} {Micromachines}\ }\textbf {\bibinfo {volume}
  {7}} (\bibinfo {year} {2016})}\BibitemShut {NoStop}%
\bibitem [{\citenamefont {Han}\ \emph {et~al.}(2016)\citenamefont {Han},
  \citenamefont {Zou},\ and\ \citenamefont {Tang}}]{Tang2016}%
  \BibitemOpen
  \bibfield  {author} {\bibinfo {author} {\bibfnamefont {Xu}~\bibnamefont
  {Han}}, \bibinfo {author} {\bibfnamefont {Chang-Ling}\ \bibnamefont {Zou}}, \
  and\ \bibinfo {author} {\bibfnamefont {Hong~X.}\ \bibnamefont {Tang}},\
  }\bibfield  {title} {\enquote {\bibinfo {title} {Multimode strong coupling in
  superconducting cavity piezoelectromechanics},}\ }\href@noop {} {\bibfield
  {journal} {\bibinfo  {journal} {Phys. Rev. Lett.}\ }\textbf {\bibinfo
  {volume} {117}},\ \bibinfo {pages} {123603} (\bibinfo {year}
  {2016})}\BibitemShut {NoStop}%
\bibitem [{\citenamefont {Chu}\ \emph {et~al.}(2017)\citenamefont {Chu},
  \citenamefont {Kharel}, \citenamefont {Renninger}, \citenamefont {Burkhart},
  \citenamefont {Frunzio}, \citenamefont {Rakich},\ and\ \citenamefont
  {Schoelkopf}}]{SchoelkopfHBAR2017}%
  \BibitemOpen
  \bibfield  {author} {\bibinfo {author} {\bibfnamefont {Yiwen}\ \bibnamefont
  {Chu}}, \bibinfo {author} {\bibfnamefont {Prashanta}\ \bibnamefont {Kharel}},
  \bibinfo {author} {\bibfnamefont {William~H.}\ \bibnamefont {Renninger}},
  \bibinfo {author} {\bibfnamefont {Luke~D.}\ \bibnamefont {Burkhart}},
  \bibinfo {author} {\bibfnamefont {Luigi}\ \bibnamefont {Frunzio}}, \bibinfo
  {author} {\bibfnamefont {Peter~T.}\ \bibnamefont {Rakich}}, \ and\ \bibinfo
  {author} {\bibfnamefont {Robert~J.}\ \bibnamefont {Schoelkopf}},\ }\bibfield
  {title} {\enquote {\bibinfo {title} {Quantum acoustics with superconducting
  qubits},}\ }\href@noop {} {\bibfield  {journal} {\bibinfo  {journal}
  {Science}\ }\textbf {\bibinfo {volume} {358}},\ \bibinfo {pages} {199--202}
  (\bibinfo {year} {2017})}\BibitemShut {NoStop}%
\bibitem [{GaN(2006)}]{GaNBAR2006}%
  \BibitemOpen
  \bibfield  {title} {\enquote {\bibinfo {title} {{GaN micromachined FBAR
  structures for microwave applications}},}\ }\href@noop {} {\bibfield
  {journal} {\bibinfo  {journal} {Superlattices and Microstructures}\ }\textbf
  {\bibinfo {volume} {40}},\ \bibinfo {pages} {426 -- 431} (\bibinfo {year}
  {2006})}\BibitemShut {NoStop}%
\bibitem [{\citenamefont {Muller}\ \emph {et~al.}(2009)\citenamefont {Muller},
  \citenamefont {Neculoiu}, \citenamefont {Konstantinidis}, \citenamefont
  {Stavrinidis}, \citenamefont {Vasilache}, \citenamefont {Cismaru},
  \citenamefont {Danila}, \citenamefont {Dragoman}, \citenamefont
  {Deligeorgis},\ and\ \citenamefont {Tsagaraki}}]{GaNBAR2009}%
  \BibitemOpen
  \bibfield  {author} {\bibinfo {author} {\bibfnamefont {A.}~\bibnamefont
  {Muller}}, \bibinfo {author} {\bibfnamefont {D.}~\bibnamefont {Neculoiu}},
  \bibinfo {author} {\bibfnamefont {G.}~\bibnamefont {Konstantinidis}},
  \bibinfo {author} {\bibfnamefont {A.}~\bibnamefont {Stavrinidis}}, \bibinfo
  {author} {\bibfnamefont {D.}~\bibnamefont {Vasilache}}, \bibinfo {author}
  {\bibfnamefont {A.}~\bibnamefont {Cismaru}}, \bibinfo {author} {\bibfnamefont
  {M.}~\bibnamefont {Danila}}, \bibinfo {author} {\bibfnamefont
  {M.}~\bibnamefont {Dragoman}}, \bibinfo {author} {\bibfnamefont
  {G.}~\bibnamefont {Deligeorgis}}, \ and\ \bibinfo {author} {\bibfnamefont
  {K.}~\bibnamefont {Tsagaraki}},\ }\bibfield  {title} {\enquote {\bibinfo
  {title} {{6.3-GHz Film Bulk Acoustic Resonator Structures Based on a Gallium
  Nitride/Silicon Thin Membrane}},}\ }\href@noop {} {\bibfield  {journal}
  {\bibinfo  {journal} {IEEE Electron Device Letters}\ }\textbf {\bibinfo
  {volume} {30}},\ \bibinfo {pages} {799--801} (\bibinfo {year}
  {2009})}\BibitemShut {NoStop}%
\bibitem [{\citenamefont {Pantazis}\ \emph {et~al.}(2010)\citenamefont
  {Pantazis}, \citenamefont {Gizeli},\ and\ \citenamefont
  {Konstantinidis}}]{GaNnems2010}%
  \BibitemOpen
  \bibfield  {author} {\bibinfo {author} {\bibfnamefont {A.~K.}\ \bibnamefont
  {Pantazis}}, \bibinfo {author} {\bibfnamefont {E.}~\bibnamefont {Gizeli}}, \
  and\ \bibinfo {author} {\bibfnamefont {G.}~\bibnamefont {Konstantinidis}},\
  }\bibfield  {title} {\enquote {\bibinfo {title} {{A high frequency GaN
  Lamb-wave sensor device}},}\ }\href@noop {} {\bibfield  {journal} {\bibinfo
  {journal} {Applied Physics Letters}\ }\textbf {\bibinfo {volume} {96}},\
  \bibinfo {pages} {194103} (\bibinfo {year} {2010})}\BibitemShut {NoStop}%
\bibitem [{\citenamefont {Rais-Zadeh}\ \emph {et~al.}(2014)\citenamefont
  {Rais-Zadeh}, \citenamefont {Gokhale}, \citenamefont {Ansari}, \citenamefont
  {Faucher}, \citenamefont {Théron}, \citenamefont {Cordier},\ and\
  \citenamefont {Buchaillot}}]{GaNreview}%
  \BibitemOpen
  \bibfield  {author} {\bibinfo {author} {\bibfnamefont {M.}~\bibnamefont
  {Rais-Zadeh}}, \bibinfo {author} {\bibfnamefont {V.~J.}\ \bibnamefont
  {Gokhale}}, \bibinfo {author} {\bibfnamefont {A.}~\bibnamefont {Ansari}},
  \bibinfo {author} {\bibfnamefont {M.}~\bibnamefont {Faucher}}, \bibinfo
  {author} {\bibfnamefont {D.}~\bibnamefont {Théron}}, \bibinfo {author}
  {\bibfnamefont {Y.}~\bibnamefont {Cordier}}, \ and\ \bibinfo {author}
  {\bibfnamefont {L.}~\bibnamefont {Buchaillot}},\ }\bibfield  {title}
  {\enquote {\bibinfo {title} {Gallium nitride as an electromechanical
  material},}\ }\href@noop {} {\bibfield  {journal} {\bibinfo  {journal}
  {Journal of Microelectromechanical Systems}\ }\textbf {\bibinfo {volume}
  {23}},\ \bibinfo {pages} {1252--1271} (\bibinfo {year} {2014})}\BibitemShut
  {NoStop}%
\bibitem [{\citenamefont {Hofheinz}\ \emph {et~al.}(2009)\citenamefont
  {Hofheinz}, \citenamefont {Wang}, \citenamefont {Ansmann}, \citenamefont
  {Bialczak}, \citenamefont {Lucero}, \citenamefont {Neeley}, \citenamefont
  {O'Connell}, \citenamefont {Sank}, \citenamefont {Wenner}, \citenamefont
  {Martinis},\ and\ \citenamefont {Cleland}}]{Martinis2009}%
  \BibitemOpen
  \bibfield  {author} {\bibinfo {author} {\bibfnamefont {Max}\ \bibnamefont
  {Hofheinz}}, \bibinfo {author} {\bibfnamefont {H.}~\bibnamefont {Wang}},
  \bibinfo {author} {\bibfnamefont {M.}~\bibnamefont {Ansmann}}, \bibinfo
  {author} {\bibfnamefont {Radoslaw~C.}\ \bibnamefont {Bialczak}}, \bibinfo
  {author} {\bibfnamefont {Erik}\ \bibnamefont {Lucero}}, \bibinfo {author}
  {\bibfnamefont {M.}~\bibnamefont {Neeley}}, \bibinfo {author} {\bibfnamefont
  {A.~D.}\ \bibnamefont {O'Connell}}, \bibinfo {author} {\bibfnamefont
  {D.}~\bibnamefont {Sank}}, \bibinfo {author} {\bibfnamefont {J.}~\bibnamefont
  {Wenner}}, \bibinfo {author} {\bibfnamefont {John~M.}\ \bibnamefont
  {Martinis}}, \ and\ \bibinfo {author} {\bibfnamefont {A.~N.}\ \bibnamefont
  {Cleland}},\ }\bibfield  {title} {\enquote {\bibinfo {title} {Synthesizing
  arbitrary quantum states in a superconducting resonator},}\ }\href@noop {}
  {\bibfield  {journal} {\bibinfo  {journal} {Nature}\ }\textbf {\bibinfo
  {volume} {459}},\ \bibinfo {pages} {546} (\bibinfo {year}
  {2009})}\BibitemShut {NoStop}%
\end{thebibliography}%

\end{document}